\documentclass[12pt]{iopart}
\usepackage{graphicx}
\usepackage{verbatim}
\def\ii{{\'{\i}}}

\font\af=msbm12

\def\CO{{\cal O}}

\def\CN{{\cal N}}

\def\be{\begin{eqnarray}}    
\def\ee{\end{eqnarray}}
\def\Dsl{\,\raise.05ex\hbox{/}\mkern-9.5mu D}
\def\mbox#1#2{\vcenter{\hrule \hbox{\vrule height#2in 
\kern#1in \vrule} \hrule}} 
\def\boxeqn#1{\vcenter{\vbox{  \hrule height2pt \hbox{\vrule
width 2pt \kern3pt\vbox{\kern3pt
\hbox{${\displaystyle #1}$}\kern3pt}\kern3pt\vrule width 2pt}\hrule height2pt}}}

\def\back{{{\raise.4em\hbox{$\, _\backslash\,$}}}}

 2

\font\blackboard=msbm10 \font\blackboards=msbm7
\font\blackboardss=msbm5
\newfam\black
\textfont\black=\blackboard
\scriptfont\black=\blackboards
\scriptscriptfont\black=\blackboardss

\def\frac#1#2{{#1\over #2}}

\def\big R{{\hbox{{\bigfield R}}}}
\def\bbig R{{\hbox{{\bbigfield R}}}}

\font\af=msbm10

\def\Z{\hbox{\af Z}}

\mathchardef\imath="717B
\def\inbar{\,\vrule height1.5ex width.4pt depth0pt}
\def\IB{\relax{\rm I\kern-.18em B}}
\def\IC{\relax\hbox{$\inbar\kern-.3em{\rm C}$}}
\def\ID{\relax{\rm I\kern-.18em D}}
\def\IE{\relax{\rm I\kern-.18em E}}
\def\IF{\relax{\rm I\kern-.18em F}}
\def\IG{\relax\hbox{$\inbar\kern-.3em{\rm G}$}}
\def\IH{\relax{\rm I\kern-.18em H}}
\def\II{\relax{\rm I\kern-.18em I}}
\def\IK{\relax{\rm I\kern-.18em K}}
\def\IL{\relax{\rm I\kern-.18em L}}
\def\IM{\relax{\rm I\kern-.18em M}}
\def\IN{\relax{\rm I\kern-.18em N}}
\def\IO{\relax\hbox{$\inbar\kern-.3em{\rm O}$}}
\def\IP{\relax{\rm I\kern-.18em P}}
\def\IQ{\relax\hbox{$\inbar\kern-.3em{\rm Q}$}}
\def\IR{\relax{\rm I\kern-.18em R}}
\font\cmss=cmss10 \font\cmsss=cmss10 at 10truept
\def\IZ{\relax\ifmmode\mathchoice
{\hbox{\cmss Z\kern-.4em Z}}{\hbox{\cmss Z\kern-.4em Z}}
{\lower.9pt\hbox{\cmsss Z\kern-.36em Z}}
{\lower1.2pt\hbox{\cmsss Z\kern-.36em Z}}\else{\cmss Z\kern-.4em Z}\fi}
\def\IGa{\relax\hbox{${\rm I}\kern-.18em\Gamma$}}
\def\IPi{\relax\hbox{${\rm I}\kern-.18em\Pi$}}
\def\ITh{\relax\hbox{$\inbar\kern-.3em\Theta$}}
\def\IOm{\relax\hbox{$\inbar\kern-3.00pt\Omega$}}

\def\CO{{\cal O}}

\def\CH{{\cal H}}

\def\CN{{\cal N}}

\begin{document}

\title{Vacuum Structure and  Boundary Renormalization Group}

\author{M. Asorey and  J. M.  Mu\~noz-Casta\~neda}

\address{Departamento de F\'\i sica Te\'orica. Facultad de Ciencias.
Universidad de Zaragoza, 50009 Zaragoza.Spain}
\ead{asorey@unizar.es}
\begin{abstract}

The  vacuum  structure is probed by   boundary conditions.
The behaviour of   thermodynamical quantities like free energy, boundary
entropy and entanglement entropy under the boundary renormalization group flow are
analysed in 2D conformal field theories. The results show that whereas vacuum energy
and boundary entropy turn out to be very sensitive to  boundary conditions,  the vacuum 
entanglement entropy  is independent of boundary properties when the boundary
of the entanglement domain does not overlap the boundary of the physical space.
In all cases the second law of thermodynamics holds along the boundary renormalization 
group flow.

\end{abstract}


\maketitle

\section{Introduction}

The vacuum of a quantum field theory has a very rich structure which in fact encodes
the complete information about  the whole quantum theory. 
Then, the response of the vacuum  to an external perturbation is a very rich source
of information not only about its own structure but also about the very nature of the 
quantum field theory.  A particularly interesting perturbation is introduced by the 
confinement of  the quantum  system to a finite space volume. In such a case, the effect 
of boundary conditions into the structure of the vacuum  gives rise to interesting 
phenomena like spontaneous symmetry breaking and other  low energy phenomena.

It is well known that the vacuum energy is very sensitive to the choice of boundary conditions.
This  dependence  generates interesting physical phenomena, like changes of sign in the
Casimir energy. 
At finite temperature  boundary conditions do not  only affect  the vacuum but also
thermodynamics quantities like the free energy and entropy. In recent years some new
quantities have been introduced to measure the entanglement structure of the quantum
vacuum \cite{sorkin} (entanglement entropy) and the structure of boundary states \cite{cardy}
(boundary entropy). The evolution of the new types of entropy seems to confirm the validity of the second law of
thermodynamics beyond the natural realm of thermodynamics. One way of testing the monotone
behaviour of the new entropies   in quantum field theory is to introduce an external
perturbation and to analyse their evolution.   In this paper we analyse the behaviour of
those quantities under the  change of
boundary conditions. In particular, we  consider the effect of such a perturbations
on a conformal invariant theory and we analyse the behaviour of the thermodynamical
quantities under the induced boundary renormalization group flow.

\section{Conformal invariance and Boundary Renormalization Group Flow}

Let us consider   a single  real massless scalar field defined in $[0,L]$. From a classical
viewpoint there is a large class of boundary conditions which can be imposed to the fields. 
However, in the quantum theory  unitarity and causality  impose severe constraints 
on the boundary behaviour of quantum fields. In particular,  the Hamiltonian with density
\be
\CH=-\frac12\,\frac{\delta^2}{\delta \phi^2} +\frac12\, \phi\, \sqrt{-\Delta}\,\phi
\label{cero}
\ee
has to be selfadjoint, which requires that the boundary conditions have 
to preserve the positivity of the Laplace-Beltrami
operator $-\Delta$. The set of boundary conditions which are compatible with unitarity and causality 
span a four-dimensional manifold which contains  the following two families of boundary conditions 
which parametrise  two complete charts of the corresponding
four-dimensional manifold \cite{aim}

a) Mixed boundary conditions
\be
L\, \varphi'(0)= -a\, \varphi(0) - b\,\varphi(L); \quad L\, \varphi'(L)= b\, \varphi(0) + d\,\varphi(L)
\label{uno}
\ee
with $a d-b^2>0$ and   $a, d \geq 0$, 
which interpolate 
  between Neumann ($a=b=d=0$) and Dirichlet ($a=b=d=\infty$)   boundary conditions; and 

b) Closed boundary conditions
\be
 \varphi(L)= \alpha\, \varphi(0) + \beta L\,\varphi '(0); \quad L\, \varphi '(L)= \gamma\, \varphi(0) + \delta L\, \varphi '(0)
\label{dos}
\ee
with $\alpha \delta+\gamma\beta=1$ and   $\alpha\gamma\geq 0,\, \beta \delta\geq 0$, 
which include  quasi-periodic boundary conditions ($\beta=\gamma=0$) that interpolate  between periodic 
 ($ \alpha=\delta =1, \beta=\gamma=0$)  and antiperiodic ($\alpha=\delta =-1, \beta=\gamma=0$)  boundary
conditions.  Notice that Zaremba boundary conditions
$$\varphi(0)=0, \varphi'(L)=0$$
can be considered  either  as mixed boundary conditions with $a=\infty, b=d=0$ or  as
 quasi-periodic boundary conditions
with $\alpha=\infty,  \gamma=\delta=0$.

Although the theory is massless, conformal invariance may be broken by the effect of boundary conditions \cite{cardy}.
The only boundary conditions which preserve conformal invariance are Neumann, Dirichlet and quasiperiodic
boundary conditions \cite{agm}. All other boundary conditions are not invariant under scale transformations and generate
a non-trivial renormalization group flow \cite{agm2}. Because the fields are non-interacting, this flow  is simply given
by
\begin{eqnarray}
\partial_t a&=-a, \qquad\quad\, 
\partial_t b=-b \qquad
\partial_t d=-d\\
\partial_t \alpha &=\partial_t \delta=0 \qquad
\partial_t \beta=\beta\qquad\,
\partial_t \gamma=-\gamma\qquad
\end{eqnarray}
where the renormalization group parameter $t$  is defined by $ L=L_0\, {\mathrm{e}}^t.$

The fixed points of this flow correspond to conformal invariant theories 
$a=b=d=0$ (Neumann), $\beta=\gamma=0$ (quasi-periodic) and $a=b=d=\infty$   (Dirichlet).
Any  boundary condition flows toward one of these fixed points.

Mixed boundary conditions flow with the boundary renormalization group  from   Dirichlet ({UV}) toward 
 Neumann ({IR}) conditions.
Critical exponents can be identified with the eigenvalues of the renormalization matrix  at the
 fixed points:   all  critical exponents are either $1$ or $-1$ which in particular shows that there are not
cyclic orbits \cite{agm}.

The most stable fixed point corresponds to Neumann  boundary conditions because
all its critical exponents are $+1$. The most unstable is that of   Dirichlet 
boundary conditions since  all critical exponents $-1$. 
 Quasi-periodic  fixed points present relevant and irrelevant
perturbations with critical exponents $\pm 1$, respectively. Negative critical exponents 
point out the instabilities. Implications of these results
for  string theory are well known \cite{agm2}.

\section{Vacuum Energy and Free Energy}

The infrared properties of quantum field theory are very sensitive to boundary
conditions \cite{karpacz}. In particular, the physical properties of the quantum 
vacuum, free energy  and  vacuum energy exhibit  a very strong dependence on 
the type of boundary conditions. 

In two dimensions there is an infrared problem which makes the analysis more subtle.
The consistency of the quantum theory is not guaranteed due to the the well known  infrared problems of massless
free bosons which prevent the existence of Goldstone phenomena like spontaneous  breakdown 
of continuous rigid symmetries \cite{Coleman}.
 The problem arises because the two point function is not positive which implies that the fundamental 
property of  Osterwalder-Schrader reflection positivity is not satisfied pointing out the inconsistency of the theory. One 
intimately related property is the odd normalization properties of the vacuum state
\be
\Psi(\phi)= \CN \ {\rm e}^{\displaystyle -\frac12 (\phi, \sqrt{-\Delta} \phi)}
\ee 
The problem persists even in finite volumes for any choice of boundary conditions and even
becomes even more dramatic  because then free  massless bosons  can 
have zero modes making the vacuum state not normalizable. One way of solving all these problems is to consider
a compactification of the scalar field $\Phi= {\rm e}^{i \phi/R}$ to a circle of unit radius. In that case the correlators
of the compactified field $\Phi$ satisfy the reflection positivity requirement and the ground state becomes normalizable
even in the zero modes sector.

The existence of the zero modes can be partially solved by the choice of boundary conditions. In fact, the
Laplace-Beltrami operator $\Delta$ 
present zero modes  only for 
the mixed conditions with $a=-b=d$, Neumann or closed boundary conditions with $\alpha=\delta=\beta=1, \gamma=0$  
The only conformally invariant conditions with zero modes are periodic and Neumann boundary conditions, i.e
the boundary conditions of the closed and open strings.

The free energy  of the system at finite temperature $1/T$ 
with the boundary conditions (2)(3)  has 
the following asymptotic expansion for large volume and low temperature $0<L<<T$ \cite{cardy1, Cardy},
\be
 {  -\hbox{Log Z}}={ f_B}\, L T  + f_b \, T + C \, \frac{T}{L}
 + { \gamma }+ \CO(1/T)
\ee
where   $f_B$ is the bulk free-energy density, $f_b$ the  
boundary energy, and  $C/L$ the Casimir energy. 
The bulk and boundary terms are UV divergent. The bulk free energy density $f_B$ 
corresponds to the infinite volume limit of free energy density and, thus, 
in any  regularization it does not depend on boundary conditions. On the
contrary the regularized boundary energy $f_b$ and 
the Casimir energy are highly dependent on boundary conditions. 
For instance,  $f_b$ is non-vanishing for Dirichlet or Neumann boundary 
conditions whereas  $f_b=0$  for periodic boundary conditions,  and
the Casimir energy  \cite{gift}
\be
 C=
\frac{\pi}{12}-\pi\, {\min_{n\in\Z}}\left[{ \frac1{\pi}\arctan \alpha}+n+
\frac1{4}\right]^2
\label{qp}
\ee
is $\alpha$-dependent  for  quasi-periodic boundary conditions. 
The values and signs of this finite size contribution to the energy are very different
for periodic ($\alpha=1,C=-\pi/6$), antiperiodic ($\alpha=-1,C=\pi/12$) and Zaremba  
($\alpha=\infty ,C=\pi/48$)  \cite{bfsv}-\cite{klp}.

For mixed boundary conditions, the Casimir energy  interpolates between the values 
$C=-\pi/24$ for Neumann ($a=b=d=0$) and
$C=-\pi/24$ for Dirichlet ($a=d=\infty$) boundary conditions \cite{agm}. Now,  since 
$C> -\pi/24$ for generic Robin boundary conditions with $0 < a=d<\infty$ and  $b=0$ \cite {aa,  agm},
 the behaviour of  Casimir energy is
not monotone along the boundary renormalization group trajectories.

\section{Boundary Entropy}

There is another asymptotic limit  of the free energy for low temperature and large volume  $0<T<<L$ of the form
\be
 {  -\hbox{Log Z}}={ f_B}\, L T   -  \, \frac{\pi L}{6T}
 + { \lambda }+ \CO(1/L).
\ee

The  second  term corresponds to the Casimir energy for periodic
boundary conditions.

There is a similar expansion for the entropy
\be
\hbox{S}= (1-T\partial_T)\log \hbox{Z}= \frac{\hbox{ }\pi L} {3T} -
\, { \lambda }+ \CO(1/L).
\ee
and the second term $\lambda=-s_b + \lambda_0$ can be
split in two pieces: one universal (independent of the boundary condition of the
fields) and another one $s_b$ which is know as boundary entropy \cite{cardy,affleck}.
This entropy $s_b=\log g$ can be formally associated with the number  of boundary states
$g$ \cite{cardy}  but in many cases $g$ is not integer  and does not
correspond to a mere counting of boundary states \cite{affleck}. It has been conjectured
that  the quantities  $g$ and $s_b$ evolve with the boundary renormalization group flow 
 in a non-increasing way \cite{affleck} 
$$s_{ _{UV}}-s_{ _{IR}}\geq 0,\quad g_{ _{UV}}-g_{_{IR}}\geq 0 $$
as it corresponds to any other type of entropy according to the second law of 
thermodynamics \cite{affleck}--\cite{friedan}. This conjecture is known as $g$-theorem and
has been verified in many cases although not yet proved \cite{affleck2}. 

Let us analyse what is the behaviour of $g$  and the boundary entropy $s_b$ along
the boundary renormalization group flow.

\def\ladrillo{\,}
\def\lubrown{\,}

 The boundary entropy can easily be computed for quasi-periodic boundary
conditions. The result turns out to be \cite{agm7}

$$s_{b}= \log |\alpha-1|-\frac12\log\left(\frac12+\frac{\alpha^2}{2}\right) ;\quad g=
\frac{\sqrt{1+\alpha^2}}{\sqrt{2}\,|\alpha-1|} $$
{for } $\alpha\neq 1$. Although
 the quasiperiodic boundary conditions describe a curve of fixed points under the
boundary renormalization group flow, and the $g$-theorem does not impose any requirement
on the behaviour of their boundary entropy, it turns out to be monotonally increasing for $\alpha<1$ and 
monotonally decreasing for $\alpha>1$, with a vanishing value at the singular 
point  of periodic boundary conditions $\alpha=1$. 
In the particular case of
Zaremba boundary conditions the value $g=1/\sqrt{2}$ agrees with the decreasing
values of the flow intepolating from Dirichlet to Neumann boundary conditions
$$ g_D=\frac{1}{{2R}} > g_Z=\frac1{\sqrt{2}}> g_N={R},$$
which is in agreement with the conjectured  g-theorem for $R< 1/\sqrt{2}$. For free massless scalars,
 the boundary entropy presents  a  behaviour under the boundary 
renormalization group similar to that of  the central charge or the
standard bulk entropy.

\section{Entanglement Entropy}

Another type of entropy that can be associated to the vacuum state $\psi_0$ is the entanglement 
entropy which to measures its degree of entanglement. The entanglement entropy is defined as the entropy
of the mixed state generated by  integrating out
the fluctuating modes of the vacuum state $\Psi_0$  in a bounded domain $(L/2-l/2, L/2+l/2)$ of the physical space
$(0,L)$ \cite{sorkin},  i.e.
\be
{\rho_l}=\int_{L/2-l/2}^{L/2+l/2}  \Psi_0^\ast \Psi_0.
\ee
The entropy  of this  state  $ S_l= -Tr \,\rho_l \log \rho_l$, is  ultraviolet 
divergent \cite{srednicki}--\cite{kabat}, but once regularised scales  logarithmically  
with the length $l$ of the interval of integration  
and the  ultraviolet cut-off $\epsilon$
 introduced to split apart  the domain $(L/2-l/2, L/2+l/2)$ and its complement 
$(0,L/2-l/2-\epsilon)\cup (0,L/2+l/2+\epsilon,L)$
\be
 S_l=\frac1{3} \log\frac{ l} { \epsilon} +\gamma(\epsilon),
\label{eone}
\ee
The coefficient $c_1=1/3$ 
of the logarithmic term in (12) is universal and does coincide with one third of the central charge of 
the corresponding conformal field theory. 
It is remarkable that  coefficient $c_1=1/3$ is also absolutely independent of the choice of boundary condition
in $(0,L)$. This can be easily understood as a consequence of the  fact that the entanglement
entropy is rather associated to the  behaviour at the interface between $(L/2-l/2, L/2+l/2)$ and its complement 
$(0,L/2-l/2)\cup (L/2+l/2,L)$  which does not depend on the choice of boundary conditions at the edge 
of the physical space.  The finite part $\gamma(\epsilon)$ is highly dependent  on the ultraviolet regularization
method. If the region where the fluctuations  are integrated out 
reaches the boundary itself, e.g. for  $(0, l)$, the entropy has the same
asymptotic behaviour \cite{calcar}
\be
 S_l=\frac1{6} \log\frac{ l} { \epsilon} + \log g+ \frac12  \gamma(\epsilon),
\label{eon}
\ee
but with a different coefficient $c_1=1/6$ for the asymptotic logarithmic term 
and a different finite term  which is also dependent on
the boundary condition and related to the boundary entropy.
The behaviour of this quantity along the boundary renormalization group flow is  then
 monotonally similar to that of the boundary entropy.
The extra $1/2$ factor  in the coefficient  of the logarithmic term can be understood  by the change on 
the number of boundary points  of the entanglement domain.
This  coefficient is in fact proportional 
to the number of
 connected components of that domain \cite{calcar}, which in this case is reduced  from two to one. The same topological
behaviour is exhibited by  the  coefficient of a similar  logarithmic term that appears in the asymptotic
expansion of the  entanglement entropy in  2+1 dimensions, and
is proportional to the Euler number of the entanglement domain \cite{Fradkin}. 

\section{Conclusions}

Zamolodchikov  introduced with his c-theorem a very interesting characterisation 
of the loss of information associated to the renormalization group flow
in 2-D quantum field theories \cite{zamol}.  This behaviour of Zamolodchikov c-function
 is very reminiscent of that of the standard bulk entropy  along such RG  flow which is
the kernel of  the second law of thermodynamics.

The Zamolodchikov c-function is associated with the central charge of
the conformal anomaly in conformally invariant field theories. Now,  for 	theories 
defined in finite volumes with boundaries the central charge also governs the
behaviour of the finite size corrections to the free energy. 
However, our results show that  the finite size corrections do not behave 
monotonally along the boundary renormalization group flow. The renormalised trajectory
which flows from the Dirichlet fixed point to the Neumann fixed point  describes a
family of systems whose free energy first increases   and then decreases 
along the same trajectory.

However, the new concepts of entropy, boundary entropy and entanglement
entropy, behave along the same trajectories as prescribes  the
second law of thermodynamics. In all  cases that we analysed the  boundary entropy
decreases along  the boundary  renormalization group flow, although  for quasi-periodic boundary
conditions the boundary entropy remains constant.

The behaviour of the entanglement entropy is completely different. It is completely
independent of the type of boundary conditions if the domain where the
quantum fluctuations of the fields  are integrated out does not reach the boundary
of the space. However, when the domain where the  quantum
fluctuations  have been averaged out reaches the
boundary, the entanglement entropy becomes dependent on the boundary conditions,
and behaves as prescribes  the second law of thermodynamics, decreasing along
the boundary renormalization group flow as the boundary entropy does.

Although the analogy between the behaviours of both new types of entropy 
along the renormalization group flow and that of the standard entropy under  time evolution 
involved  the  second law of thermodynamics
might appear as incidentally due to similar mechanisms of information loss and irreversibility,
both behaviours have, in fact, a closer relation from a physical viewpoint.
The renormalization group flow of the quantum vacuum or the  finite temperature 
canonical density matrix can also be though  as a dynamical adiabatic evolution under 
the one-parametric family of Hamiltonians connected by scale transformations.
Thus, according to the thermodynamical principles,
the behaviour of both states under the renormalization group flow 
 must be compatible with the second law.
On the other hand this kind of physical time evolution is quite natural from the 
cosmological point of view, because any physical system is coupled to the 
underlying background space metric which is continuously expanding towards 
the infrared along the cosmic time.

The same analysis can also be  performed in the presence of bulk interactions, e.g
a $\lambda \phi^4$ term. The renormalization group flow changes by the effect of
the perturbation. Some new fixed points appear and some other disappear.
In particular, new  self-interacting conformal field theories can appear without the
infrared problems associated to the free field theories \cite{zamol2}.  The analysis of the behaviour
of boundary entropy and entanglement entropy in those cases is a challenging  open 
problem.

\section*{Acknowledgements}
We thank M. Aguado, 
J.G. Esteve, J.I. Latorre and  G. Marmo, for interesting discussions. 
This work is partially supported by CICYT (grant FPA2006-2315)
and DGIID-DGA (grant2007-E24/2).

\bigskip


\vspace{40pt}
\section*{References}
\begin{thebibliography}{10}


\bibitem{sorkin} L. Bombelli, R.K. Koul, J. Lee and R. Sorkin,
Phys. Rev. {\bf D 34} (1986) 373

\bibitem{cardy} J. Cardy, Nucl. Phys. {\bf B324} (1989) 581 



\bibitem{aim}  M. Asorey, A. Ibort and G. Marmo,
 Int. J. Mod. Phys. {\bf A 20} (2005) 1001


\bibitem{agm} M. Asorey, D. 
Garc\ii a-Alvarez and J. M. Mu\~noz-Casta\~neda, J. Phys. {\bf  A 39} (2006) 6127-6136 
\bibitem{agm2} M.  Asorey, D. 
Garc\ii a-Alvarez and J. M. Mu\~noz-Casta\~neda, J. Phys. {\bf A 40} (2007) 6767-6775

\bibitem{karpacz}
M. Asorey, J. Geom. Phys.  {\bf 11} (1993)94



\bibitem{Coleman}
S. Coleman, Commun.
Math. Phys. {\bf 31}  (1973) 259

\bibitem{cardy1} J. Cardy,  Nucl. Phys.
{\bf B 240}  (1984) 514


\bibitem{Cardy}
J. Cardy,  In {\it Encyclopedia of Mathematical Physics}, Eds.
J.-P. Fran\c coise, G. L. Naber and
T. S. Tsun, Academic Press (2006)


 \bibitem{gift}
 M. Asorey, unpublished (1994)

\bibitem{bfsv}
 M. Bordag, H. Falomir, E. M. Santangelo and D. V. Vassilevich,
Phys.Rev. {\bf D65} (2002) 064032
\bibitem{saharian}
A. A. Saharian,
Phys. Rev. {\bf  D69} (2004) 085005
\bibitem{bordag}M. Bordag, U. Mohideen, V. M. Mostepanenko,Phys. Rep. 
{\bf 353} (2002) 1-205
\bibitem{aa}
A. Romeo and A. A. Saharian,
J. Phys. A35 (2002) 1297-1320



\bibitem{m2} J. M\"uller and W. M\"uller, Duke Math. J., {\bf 133} (2006), 259

\bibitem{klp} K. Kirsten, P. Loya and J. Park, Ann. Phys.  (NY), {\bf 321} (2006)1814



\bibitem{affleck} I. Affleck and A.W.W. Ludwig, Phys. Rev. Lett. {\bf 67} (1991) 161 
.

\bibitem{friedan} D. Friedan and  A. Konechny,  Phys. Rev. Lett. {\bf 93}  (2004) 030402.

\bibitem{affleck2} I. Affleck and A. W. W. Ludwig, Phys. Rev. {\bf B48} (1993) 7297


\bibitem{agm7} M. Asorey and J. M. Mu\~noz-Casta\~neda, In preparation. 



 \bibitem{srednicki} M. Srednicki, Phys. Rev. Lett. {\bf 71} (1993) 666

\bibitem{callan} C. G. Callan and F. Wilczek,  Phys. Lett. {\bf B333}(1994)55
\bibitem{dowker} J. S. Dowker, Class. Quant. Grav. {\bf 11} (1994) 55
\bibitem{kabat}  D. Kabat and M. J. Strassler, Phys. Lett. {\bf B329}(1994)46

\bibitem{calcar}
P. Calabrese and J. Cardy, J. Stat. Mech. {\bf 0406} (2004) 002

\bibitem{Fradkin} E. Fradkin and  J. E. Moore,  Phys. Rev. Lett. {\bf 97} (2006) 050404


\bibitem{zamol} A. B. Zamolodchikov, JETP Lett. {\bf 43} (1986) 730-732
\bibitem{zamol2} A. B. Zamolodchikov, Sov. J. Nucl. Phys. {\bf 44} (1986)529

\end{thebibliography}
\end{document}